# About stability of plasma in a magnetic field

**Vyacheslav Buts**[1,2,3]

[1] National Science Center "Kharkov Institute of Physics and Technology", Kharkov, Ukraine.
[2] V.N. Karazin Kharkov National University, Kharkov, Ukraine.
[3] Institute of Radio Astronomy of NAS of Ukraine, Kharkov.
\* Correspondence: E-mail: vbuts@kipt.kharkov.ua

**Abstract:** The dynamics of the oscillator system is investigated. The conditions under which this dynamics becomes unstable are determined. In particular, it is shown that plasma in constant magnetic field becomes unstable if its density exceeds a certain critical value. The role of chaotic regimes on confinement of particles (oscillators) in confining potentials is studied.

## 1. Introduction

Plasma, located in an external magnetic field, is a key component in the program of controlled thermonuclear fusion. The features of its confinement in various magnetic configurations are studied in detail. Numerous hydrodynamic and kinetic instabilities have been investigated (see, for example, [1-3]). It has been shown [4-6] that at plasma densities $> 10^{14}$ cm$^{-3}$ there are problems with its retention. This phenomenon has even acquired its own name "density limit". A phenomenological criterion for the plasma density limit, the Greenwald limit, is obtained. However, the physical mechanisms responsible for existence of restrictions on the density of confined plasma are still unknown. For years there is a search for ways to overcome this limit. In [7], as an experimental success, is reported that at the Alcator C-Mod tokamak a plasma density of $1.5 \times 10^{14}$ cm$^{-3}$ was obtained, i.e. one and a half times the calculated limit.

It is possible that the mechanisms described in this paper will to some extent allow a deeper understanding of the cause of the "density limit". Two types of instabilities are considered here. The first is related to the fact that under certain conditions the dynamics of a system of linear and nonlinear coupled oscillators ceases to be oscillatory, and becomes unstable. Such a system throws out "extra" particles from its ensemble. Such dynamics takes place with certain characteristics of the connection between oscillators, and depends on the number of oscillators in the ensemble. Note that the well-known methods for analyzing the dynamics of systems consisting of a large number of oscillators (e. g., in [8 - 10]), formulate in one form or another conditions under which the dynamics of an ensemble of coupled oscillators remain oscillatory. Such an approach is natural, since unstable ensembles do not exist. They fall apart. Therefore, only the properties of oscillatory ensembles are studied. As a result, the question of the development of instabilities in a system of a large number of coupled oscillators has been little studied. The second type of instability is associated with the development of regimes with dynamic chaos in the dynamics of systems of nonlinear oscillators.

Below, in the second and third sections, we consider simple, but important for some applications, systems of coupled oscillators. The conditions for their instability are determined. In the fourth section, a system of nonlinear oscillators has been studied (mainly by numerical methods). It is shown that in the regime with dynamic chaos they can exist only with certain characteristics of the confining potential and characteristics of the chaotic dynamics. It is shown that even a single nonlinear oscillator in the dynamic chaos mode at sufficiently large values of the moments ceases to be held by the potential. The fifth section is auxiliary. It provides additional evidence that particle dynamics in plasma is characterized by chaotic dynamics. Finally, in the sixth section, it is shown that the plasma, which is in a magnetic field, can stably exist if its density does not exceed a certain critical value. 'Extra' particles will be ejected from the ensemble. In conclusion, the main results are formulated.

**2. Dynamics of Ensemble of linear oscillators**

Suppose we have a system with Hamiltonian:

$$H = \sum_{i=0}^{N}\left(\frac{p_i^2}{2} + \omega_0^2 \frac{q_i^2}{2}\right) + \mu \cdot q_0 \cdot \sum_{j=1}^{N} q_j \tag{1}$$

This system is a system of coupled linear oscillators. Hamiltonian (1) corresponds to the system of equations for describing the dynamics of coupled linear oscillators:

$$\ddot{q}_i + \omega_0^2 q_i = -\mu \cdot q_0$$
$$\ddot{q}_0 + \omega_0^2 q_0 = -\mu \cdot \sum_{i=1}^{N} q_i \tag{2}$$

For simplicity, we consider a system in which all oscillators are connected with each other only through a zero oscillator (see fig. 1). The normal frequencies of such a system are easy to find. To do this, we will look for the solution of system (2) in the form:

$$q_i = a_i \exp(i \cdot \omega \cdot t), \quad a_i = const \tag{3}$$

Substituting this solution into (2), we obtain the dispersion equation:

$$\left(-\omega^2 + \omega_0^2\right)^2 = \mu^2 N. \tag{4}$$

Equation (4) gives the following expressions for normal frequencies:

$$\omega = \pm \omega_0 \sqrt{1 \pm \mu \cdot \sqrt{N}/\omega_0^2}. \tag{5}$$

The signs + and - in the formula (5) before the root and under the root are independent. It can be seen that even with a very small coupling coefficient, but with a large number of oscillators, one of the normal frequencies can be very small (for the case of the sign (-) under the root). If the inequality holds:

$$\mu \cdot \sqrt{N} > \omega_0^2 \tag{6}$$

then such an ensemble cannot exist. It collapses. A numerical analysis of the dynamics of the system (2) fully confirms this result. So, for example, if ten oscillators at the initial time placed randomly in the vicinity of the bottom of the potential well (see fig. 3) and the coupling coefficient is less than 0.3, then oscillations of the oscillators are limited. However, if we slightly increase the coupling coefficient ($\mu = 0.3334$), then the dynamics become unstable. Failure criterion (6) is fulfilled. The ensemble collapses (see fig.4).

The ensemble considered above is a simplest model. The ensemble of oscillators presented in figure 2 seems to be more realistic. In this ensemble, as in the previous one, all oscillators are connected with the central oscillator, and also coupled with their nearest neighbors. In addition, the frequency of the central oscillator is different from the frequency of the other oscillators. The system of equations that describes the dynamics of such an ensemble has the following form:

$$\ddot{q}_i + \omega_0^2 q_i = -\mu \cdot q_0 - \mu_1(q_{i+1} + q_{i-1})$$
$$\ddot{q}_0 + \omega_1^2 q_0 = -\mu \cdot \sum_{i=1}^{N} q_i \tag{7}$$

To find the conditions for the existence of such an ensemble, it is convenient to rewrite the system (7) in the form:

$$\ddot{Q} + \omega^2 q_i = -\mu \cdot N q_0$$
$$\ddot{q}_0 + \omega_1^2 q_0 = -\mu \cdot Q \quad (8)$$

where $\omega^2 = \omega_0^2 + 2\mu$, $Q = \sum_{i=1}^{N} q_i$.

We will seek solutions to system (8) in the form: $Q \sim \exp(i \cdot \omega \cdot t)$. Then for normal frequencies the following expression can be obtained:

$$\omega^2 = \frac{1}{2}\left(\omega_1^2 + \omega_2^2\right)\left[1 \pm \sqrt{1 - 4\left(\omega_1^2 \cdot \omega_2^2 - \mu^2 N\right)/\left(\omega_1^2 + \omega_2^2\right)^2}\right] \quad (9)$$

with $\omega_2^2 = \omega_0^2 + 2\mu$.

All the features of the dynamics of such a system are similar to the features of the previous system of oscillators. The condition for the destruction of this system of oscillators will be the condition:

$$\mu^2 N > \omega_2^2 \omega_1^2 \quad (10)$$

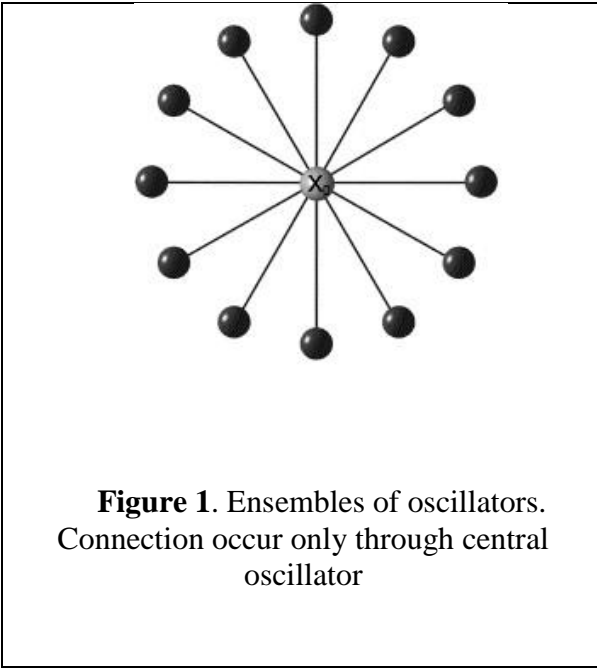

**Figure 1**. Ensembles of oscillators. Connection occur only through central oscillator

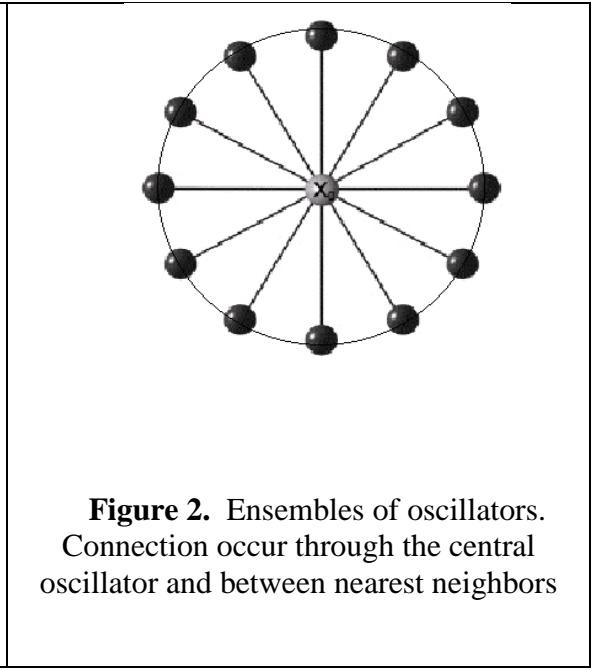

**Figure 2.** Ensembles of oscillators. Connection occur through the central oscillator and between nearest neighbors

## 3. Linear oscillators. Everyone is connected to everyone

It is of interest to consider the dynamics of the most frequently encountered system of linear oscillators, in which each oscillator is associated with all other oscillators. The system of equations that describes the dynamics of such oscillators can be written as:

$$\ddot{x}_i + \omega_0^2 x_i = -\mu \cdot \sum_{j \neq i}^{N} x_j = -\mu \cdot Q + \mu \cdot x_i \quad , \quad (11)$$

where $Q = \sum_{i=0}^{N} x_i$.

Having introduced a new function, the system of equations (11) can be rewritten in a more convenient form:

$$\ddot{Q} + \omega_0^2 Q = -\mu \cdot (N-1) Q \quad (12)$$

From equation (12) it follows that with $\mu > 0$, the dynamics of the oscillator system remains oscillatory. However, with $\mu < 0$ and large number of oscillators ($N > 1 + \omega_0^2/|\mu|$), instability appears. For example, let $N = 10$, $\omega_0 = 1$, then, if $|\mu| < 0.1$, the dynamics are regular and oscillatory. But already with $|\mu| \geq 0.2$ the ensemble is unstable, and the numerical calculations confirm these results.

## 4. Ensemble of nonlinear oscillators

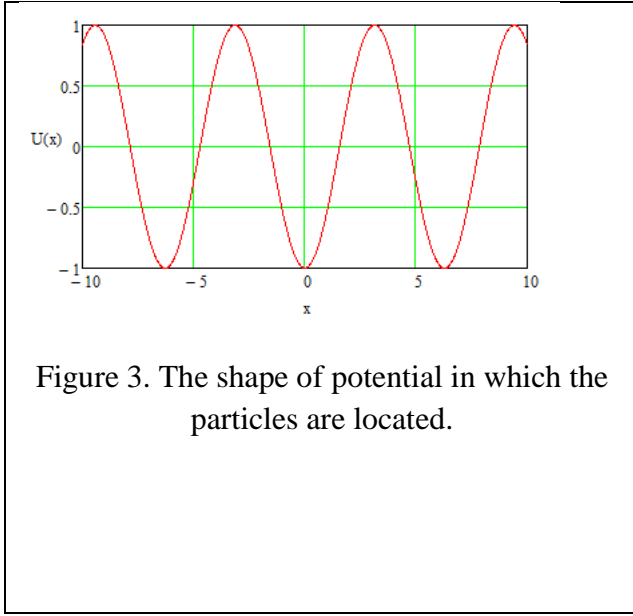

Figure 3. The shape of potential in which the particles are located.

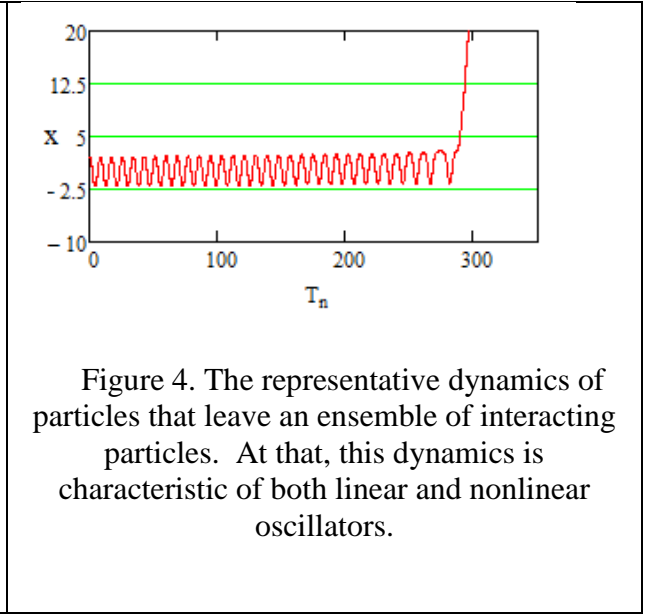

Figure 4. The representative dynamics of particles that leave an ensemble of interacting particles. At that, this dynamics is characteristic of both linear and nonlinear oscillators.

Let's write the system of differential equations, which describes the dynamics of an ensemble of mathematical pendulums, in which each oscillator is connected with any other:

$$\ddot{x}_i + \sin x_i = -\mu \cdot \sum_{j \neq i}^{N} x_j \qquad i, j \in \{0, 1, 2....N\} \tag{13}$$

The system (13) is a generalization of system (11) to the case when the oscillators are nonlinear. The system of equation (13) was investigated by numerical methods. The coupling coefficient between the oscillators was chosen small ($\mu < 10^{-3}$). The particles, at the initial time were located in the vicinity of the equilibrium state of charged particles. In the general case, during certain time interval, the particles oscillate in the potential having been captured by this potential. However, depending on the magnitude of the coupling coefficient, the initial position of the particles in the potential and on their number, some of the particles are ejected from the potential. Departure of one of the particles is shown in the figure 4. This case corresponds to the dynamics of ten oscillators that are connected by the coupling coefficient $\mu = 0.0005$. The maximum dimensionless initial velocity of one of the particles (not necessarily the one ejected) was equal to 0.4. It should be noted that the time of particle departure from the potential is very sensitive to small changes in the potential itself, the coupling coefficients, and particle position. To this one can add that the dynamics of all particles is locally unstable. The dynamic chaos is developing. If one removes the particles that flew out of the ensemble, then the dynamics of the remaining particles remain restricted.

## 5. Chaotic dynamics of the particles in plasma

The motion of particles in plasma generally obeys random dynamics. This fact is noted when describing the results of numerous theoretical and experimental works (for example, [11–12]). Strict proof of this fact and the evaluation of the criteria for the emergence of regimes with dynamic chaos are available for cyclotron resonances (e.g., [13–16]). Particle dynamics in the fields of numerous waves that are excited in a plasma must also be chaotic (in the absence of cyclotron resonances). Intuitively, ehis presentation is not in doubt. However, there is no strict evidence of this fact and the conditions for the emergence of such dynamics. Below, a simple example shows that in a fairly general case, this fact can be proved quite strictly. This can be done as follows. It is known that plasma, especially plasma in a magnetic field, has a rich spectrum of natural waves. It is possible to show that plasma particles in these, even regular fields, move chaotically. For proof, let's consider the motion of charged particles in the field of a wave packet:

$$\ddot{z} = \frac{e}{m} \sum_i E_i \sin(k_i z - \omega_i t). \tag{14}$$

To start, for the dynamics in the field of a single wave one can obtain the well-known integral from equation (14):

$$\frac{\dot{\varphi}^2}{2} - \Omega^2 \cos\varphi = H = const. \tag{15}$$

Here: $\varphi = kz - \omega t$ $\quad \Omega^2 = \frac{|e|Ek}{m\omega^2}$ $\quad \dot{\varphi} = d\varphi/d\tau; \tau = \omega t$.

Then using integral (15), the width of the nonlinear resonance is:
$$\dot{\varphi}_{max} = +2\Omega, \quad \dot{\varphi}_{min} = -2\Omega. \tag{16}$$

To determine the distance between the resonances, we note that the effective interaction of particles with the packet of waves occurs under the conditions of the Cherenkov resonance. In this case, the distance between the resonances is:

$$\Delta\dot{\varphi} = -\Delta k \left[ v_0 - \frac{\Delta\omega}{\Delta k} k_0 \right]. \tag{17}$$

At deriving (17), was taken into account that $v = v_{ph} = \omega/k$.
Using expressions (16) and (17), we find the conditions for the occurrence of a local instability:

$$K = \left(\frac{\omega}{\Delta\omega}\right) \frac{2\Omega}{\left[1 - v_g/v_{ph}\right]} = N \frac{2\Omega}{\left[1 - v_g/v_{ph}\right]}. \quad K > 1 \tag{18}$$

Here $v_g$ is the group velocity; N is the number of waves in the packet.

Analyzing formulas (17) and (18), several conclusions can be made. The first is clear (from formula (17)): if the group velocity tends to the phase velocity of the wave, then the distance between the resonances tends to zero. This means that all the waves of the packet are located on the straight line of the dispersion. In the phase space, the resonances of such waves all coincide. For particles, such resonances are practically indistinguishable, thus the dynamics should be regular. Secondly, if the group velocity of the waves tends to zero (for example, Langmuir waves in a plasma), then, we can have the inequality $1 < K \ll N$, $\Omega \ll 1$. In this case, as it was for the first time, apparently, noted in [17], the particle dynamics should be chaotic. It should be noted that the nonrelativistic dynamics of particles almost always corresponds to the case $\Omega \ll 1$.

To the result obtained, one can add additional arguments in favor of the fact that the particle dynamics in the plasma is chaotic. Except, it can be shown that the chaotic dynamics of particles can significantly facilitate their escape from the confining potential. Let, for example, particles move, as in the previous section, in the potential shown in fig. 3. The particle dynamics in such a potential is described by the equation (13). The displacement of each of these oscillators can be represented as:

$$x_i = \bar{x} + \delta_i, \tag{19}$$

where $\bar{x} = \left(\sum_{i=0}^{N} x_i\right)/N$ is the mean coordinate of the nonlinear oscillator displacement; $\delta_i$ - random deviation with normal distribution, and $\langle \delta_i \rangle = 0$.

The value $\bar{x}$ corresponds to the "center of inertia" for ensemble of oscillators. In this case, the average values over the ensemble of the function $\sin x$ can be conveniently represented as a series by moments:

$$\langle \sin(x_i) \rangle = \left[1 - \sum_{m=1}^{\infty} \frac{M_{2m}}{(2m!)}\right] \sin \bar{x}, \tag{20}$$

where $M_n = \langle (\delta)^n \rangle$ - moments.

Then an equation for description of $\bar{x}$ take the form:

$$\ddot{\bar{x}} + \left[1 - \sum_{m=1}^{\infty} \frac{M_{2m}}{(2m!)}\right] \sin \bar{x} = 0 \tag{21}$$

The region of localization of particles becomes smaller with increasing numbers and value of moments as follows from (21). This region is determined by the width of the nonlinear resonance:

$$\Delta = 4 \cdot \sqrt{\left[1 - \sum_{m=1}^{\infty} M_{2m}/(2m!)\right]}$$

As a result, the depth of the effective potential well, in which particles move, also decreases. Therefore, even small external forces easily eject particles from the capture region.

## 6. Dynamics of plasma particles in a magnetic field

Consider the motion of particles with a charge $e$ in an external magnetic field directed along the axis $z$: $\vec{H}_0 = \{0, 0, H\}$. The particles rotate around the magnetic field lines, thus move with acceleration and radiate. We will consider the ideal plasma; therefore, the Coulomb interaction of particles will not be taken into account. Let's assume that the interaction is carried out using only fields that particles emit during rotation. The electric field strength in the vicinity of another particle will be:

$$\vec{E} = -\frac{e \cdot \dot{\vec{v}}}{c^2 R} \tag{22}$$

Here $e$ - is the charge of the particle, $\dot{\vec{v}}$ - is the acceleration of the particle as it rotates, $c$ - is the velocity of light, $R$ - is the distance between particles.

The interaction of two particles will be examined first with the motion transverse relative to the magnetic field. The dynamics of one particle can be described by the following equation:

$$m\ddot{\vec{r}} = \frac{e}{c}\left[\vec{v}\vec{H}_0\right] - \frac{e^2 \dot{\vec{v}}}{c^2 R} \tag{23}$$

This vector equation is convenient to rewrite in the form of a system of equations for the velocity components of the first particle:

$$\dot{v}_{x1} = \omega_H v_{y1} - \mu \cdot \dot{v}_{x2} \tag{24}$$
$$\dot{v}_{y1} = -\omega_H \cdot v_{x1} - \mu \cdot \dot{v}_{y2},$$

where $\omega_H = \dfrac{eH}{mc}$ is the cyclotron frequency of rotation of the particle in magnetic field, and $\mu = \dfrac{e^2}{Rmc^2}$ is the influence of the second particle on the dynamics of the first particle (coupling coefficient between particles).

Differentiating the system of equations (24) and taking into account the system (24) itself, one can obtain the following system of equations:

$$(\ddot{v}_{x1} + \omega_H^2 \cdot v_{x1}) = 2\mu\omega_H^2 v_{x2}$$
$$\ddot{v}_{y1} + \omega_H^2 v_{y1} = 2\mu\omega_H^2 v_{y2} \qquad (25)$$

Similar systems of equations can be written for the dynamics of the second particle. If there are many particles, then the equations that describe the dynamics of the velocity components can be represented as:

$$\ddot{v}_k + v_k = 2\sum_{j\neq k}\mu_j v_j \qquad (26)$$

The dependent variable $v_k$ in equation (26) defines either $x$ or the $y$ component of the velocity of the $k$ particle. In addition, there has been introduced a new time $\tau = \omega_H t$. The coefficients of the connection $\mu_j$, which stand in the right side of equation (26) under the sign of the sum, differ from each other only by the distance between the particles $R_j$. Note that with a large number of oscillators ($N \gg 1$) and small coupling coefficients ($\mu_j \ll 1$), the right side of equations (26) is the same for all oscillators (for all $k$). In such a case, equations (26) can be significantly simplified:

$$\ddot{v}_k + \left(1 - 2\sum_{j=1}^{N}\mu_j\right)v_k = 0 \qquad (27)$$

and the condition of instability has the form:

$$\sum_{j=1}^{N}\mu_j > \dfrac{1}{2} \qquad (28)$$

The condition (28) can be rewritten for particle density. Indeed, the total number of interacting particles is equal to the density of particles multiplied by the volume occupied by the interacting particles ($N = nV$). The volume of a spherical layer of radius $r$ and thickness $dr$ is equal $V_{laer} = 4\pi r^2 dr$. Using this, the left side of inequality (28) can be rewritten:

$$2\sum_{j=1}^{N}\mu_j = 8\pi n\int_0^{R_{max}}\dfrac{r^2 dr}{r}\dfrac{e^2}{mc^2} = 4\pi\dfrac{e^2}{mc^2}nR_{max}^2$$

And the condition (28) will take the form:

$$n > 3\cdot 10^{11}/R_{max}^2 \qquad (29)$$

**7. Discussion and conclusion**

Let's formulate the most important results:
1. An ensemble of charged particles (electrons) in magnetic field can stably exist only if the density of these particles is less than a certain critical number that can be found from formula (29)). The "extra" particles are removed from the ensemble (see sections 2–4). The development of this instability means that plasma, for example a plasma cylinder, at densities above $10^{11}$ begins to dress with a coat of high-energy electrons. The plasma rod will be positively charged. As result there appear holding potential. The dynamics of particles in the coat is chaotic.

2. The chaotic dynamics of particles in the holding potentials can significantly weaken the efficiency of these particles confinement (according to formula (21)).
3. Above, the main attention was paid to the conditions for the destruction of the oscillatory dynamics of an oscillator system. However, stable ensembles with a large number of oscillators can also have considerable interest. Examples of such ensembles are given in Section 2. The normal frequency of the ensemble can be significantly lower than the partial frequencies of individual oscillators. This occurs when the number of the oscillators is high enough (formula (5)). Herewith, if at some point in time the main oscillator, which is connected with all other oscillators, disappears, then the oscillators begin to oscillate with their partial frequency. This feature can be used. For example, if a modulated electron beam of finite duration ( $0 < t < T$ ) is considered as the main binding oscillator, then after a time period $T$ it disappears. If the beam was modulated at the normal frequency of the ensemble, it will resonantly excite oscillations of the entire ensemble. After its disappearance, individual oscillators will oscillate at their partial frequencies. Such mechanism allows one to efficiently convert the energy of a low-frequency oscillation (the energy of a modulated beam) into the energy of high-frequency oscillations (partial frequency). This mechanism can be used to create new types of radiation sources in poorly mastered frequency ranges. For example, in the terahertz range.
4. The question arises as to which systems of oscillators could be stable, i.e. their dynamics will remain oscillatory regardless of the number of oscillators. In particular, in monographs [8, 9] the oscillator systems are considered whose frequencies, as well as coupling coefficients between oscillators, do completely depend on the characteristics of the potential in which the charged particles (oscillators) are located. Moreover, each oscillator is connected with all other oscillators, and the coupling coefficients are reciprocal - they are symmetric in their indices $\mu_{ik} = \mu_{ki}$. In this case, such oscillator systems have only oscillatory modes and instability in such ensembles does not develop. The results obtained above allow us to distinguish two groups of oscillators in which the development of instabilities is possible. The first group includes oscillator ensembles, where not all oscillators are connected with all others. The examples are the oscillator systems discussed in the second section. Such oscillator systems are easy to implement artificially. In plasma, in natural conditions, they do not exist. However, in plasma, especially in plasma held by a magnetic field, one can distinguish the second group of unstable oscillator ensembles. The connection between the oscillators in this group is determined by the fields that are excited by the oscillators themselves. Plasma is diamagnetic, so these fields impede the process that excited them. The result of this physical feature is a change in the sign of the coupling coefficients. With a sufficiently large number of oscillators, such systems, as we have seen above, become unstable.
5. The models described above are as simple as possible. In the case, when under experimental conditions, for example, the magnetic field is inhomogeneous, the partial frequencies of the oscillators become different. The efficiency of the interaction of oscillators with different frequencies is much less than the efficiency of the interaction of identical oscillators. Therefore, it can be expected that the required number of particles for the destruction of an ensemble will be greater than, for example, formula (28) determines. However, the tendency toward the appearance of normal modes, the frequency of which is much less than the partial frequencies of individual oscillators, will obviously remain. In this case, a collective low-frequency mode may appear. These modes can eject plasma particles onto the wall. Note that the analysis of such systems is difficult even by numerical methods.

Author is grateful to Professor V.S. Voitsenya, who turned attention to the problem "density limit", for useful advices and for editing the English text.

**References**

1. Nicholas A. Krall, Alvin W. Trivelpiece, Principles of Plasma Physics, McGRAW-HILL Book Company .1973, 526
2. Chen, F.F.; *Introduction to Plasma Physics and Controlled Fusion*. 1984, l,2 nded. (Plenum Press,).
3. Fusion physics. Edited by: Mitsuru Kikuchi; Karl Lackner; Minh Quang. *Trans International Atomic Energy* Agency, Vienna, 2012.
4. Greenwald, M., Density limits in toroidal plasmas, *Plasma Phys. Control. Fusion,* 2002, 44 , R27-R53.
5. Puiatti, M.E.; Scarin, P.; Spizzo, G.; et al.. High density limit in reversed field pinches. *Phys. Plasmas.* 2009, 16, 012505
6. Gates, D.A.; Delgado-Aparicio, L. Origin of Tokamak Density Limit Scalings, *Phys. Rev. Lett.* 2012, 108,165004.
7. Baek, S.G.; Wallace, G.M.; Bonoli, P.T.; Brunner, D.; Faust, I.C.; Hubbard, ; Hughes, J.W. ;. LaBombard, B.; Parker, R.R; Porkolab, M.; Shiraiwa, S.; and Wukitch, S. Observation of Efficient Lower Hybrid Current Drive at High Density in Diverted Plasmas on the Alcator C-Mod Tokamak. *Phys. Rev. Lett.* 121, 055001 – Published 3 August 2018.
8. Landau, L. D.; Lifshitz, E.M. Mechanics, *Elsevier Butterworth Heinemann,* 172.
9. Magnus, K. Oscillations. *Moscow "Mir"*. 1982, 304 (In Russia).
10. Kuklin, V.M.; Litvinov, D.N.; Sevidov, S.M.; Sporov, A.E. Simulation of synchronization of nonlinear oscillators by the external field. *East European Journal of Physics*. 2017, 4, (1), 75-84.
11. Shustin, E.G.; Isaev, N.V.; Temiryazeva, M.P.; Tarakanov, V.P.; Fedorov, Yu.V. Beam-Plasma Discharge in a Weak Magnetic Field as a Source of Plasma For a Plasma- Chemical Reactor *Kharkov. Problems of Atomic Science and Technology. Series: Plasma electronics and new acceleration methods.* 2008, № 4 (6), 169.
12. Meshkov, I.N.; Nagaitsev, S.S.; Seleznev, I.A.; Syresin, E.M. Beam – Plasma Discharge at Electron Beam Injection Into Rare Gas. Preprint 90-12. *Novosibirsk Institute of Nuclear Physics.* 1990, 630090, 15.
13. Moiseev, S.S.; Buts, V.A.; Erokhin, N.S. Peculiarities of Charged Particle Dynamics under Cyclotron Resonance Conditions. *Plasma Physics Reports*. 2016, 42, N 8, 761-768.
14. Buts, V. A.; Kuzmin, V. V.; Tolstoluzhsky, A. P. Features of the dynamics of particles and fields at cyclotron resonances. *ZhETF,* 2017, 152, 4 (10), 767-780.
15. Buts, V. A.; Lebedev A. N.; and Kurilko V. I., The Theory of Coherent Radiation by Intense Electron Beams. *Springer,* Berlin, 2006.
16. Buts, V.A. Regular and chaotic dynamics of charged particles during wave-particle interactions. *Problems of theoretical physics. Series. Problems of theoretical and mathematical physics.* 2017, Kharkiv, 2, 122-241.
17. Zaslavsky, G.M.; Chirikov, B.V.; Stochastic instability of nonlinear oscillations. *Phys. Usp. 1971*, 105, 1-42.

# ОБ УСТОЙЧИВОСТИ ПЛАЗМЫ В МАГНИТНОМ ПОЛЕ

*V.A. Buts*

*National Science Center "Kharkov Institute of Physics and Technology", Kharkov, Ukraine;*
*V.N. Karazin Kharkov National University, Kharkov, Ukraine;*
*Institute of Radio Astronomy of NAS of Ukraine, Kharkov, Ukraine*
*E-mail: vbuts@kipt.kharkov.ua; vavriv@rian.kharkov.ua*

Исследована динамика системы осцилляторов. Определены условия, при которых эта динамика становится неустойчивой. В частности, показано, что плазма в постоянном магнитном поле становится неустойчивой, если ее плотность превышает некоторую критическую величину. Изучена роль хаотических режимов на удержание частиц (осцилляторов) в удерживающих потенциалах.

**Keywords**: линейные осцилляторы, нелинейные осцилляторы, ансамбли осцилляторов, неустойчивость, динамический хаос, плазма

1. Introduction

Плазма, находящаяся во внешнем магнитном поле, является одним из ключевых элементов в программе управляемого термоядерного синтеза. Ее динамика детально изучена. Изучены многочисленные гидродинамические и кинетические неустойчивости (см., например, [1-3]). Особое внимание следует обратить на проблему удержания плазмы в установках с магнитным удержанием плазмы. Известно [4-6] , что при плотностях >$10^{14}$ см$^{-3}$ удерживать плазму практически не удается. Этот факт (феномен) приобрел даже собственное имя "density limit" . Имеется феноменологический критерий предела плотности плазмы – критерий «Greenwald limit». Однако физические механизмы, которые ответственны за существование ограничений плотности удерживаемой плазмы, до настоящего времени неизвестны. Идет борьба за преодоление этого предела. В работе [7] в качестве достижения сообщается, что на токамаке Alcator C-Mod в Plasma Science and Fusion Center была получена плотность удерживаемой плазмы 1.5x$10^{14}$ см$^{-3}$ , т.е. в полтора раза больше существующего предела. Возможно, что описанные в этой работы механизмы в какой-то степени позволят более глубоко разобраться с феноменом "density limit". Рассмотрены два типа неустойчивостей. Первый тип неустойчивостей связан с тем фактом, что при определенных условиях динамика системы линейных и нелинейных связанных осцилляторов перестает быть колебательной. Она становится неустойчивой. Такая система выбрасывает лишние частицы из своего ансамбля. Такая динамика происходит при определенных характеристиках связи между осцилляторами, а также от количества осцилляторов в ансамбле. Отметим, что известные методы анализа динамики систем, состоящих из большого числа осцилляторов (см., например, [8-10]), в той или иной форме формулируют условия, при которых динамика ансамбля связанных осцилляторов остается колебательной. Такой подход естественен, так как неустойчивые ансамбли не существуют. Они распадаются. Поэтому изучаются свойства только колебательных ансамблей. В результате вопрос о развитии неустойчивостей в системе большого числа связанных осцилляторов изучен мало (недостаточно). Второй тип неустойчивостей связан с развитием режимов с динамическим хаосом в динамике систем нелинейных осцилляторов.
    Ниже во втором и третьем разделах рассмотрены простые, но важные для некоторых приложение, системы связанных осцилляторов. Определены условия их неустойчивостей. В

четвертом разделе изучена (в основном численными методами) система нелинейных осцилляторов. Показано, что в режиме с динамическим хаосом они могут существовать только при определенных характеристиках удерживающего потенциала и характеристик хаотической динамики. Показано, что даже один нелинейный осциллятор в режиме динамического хаоса при достаточно больших величинах моментов перестает удерживаться потенциалом.

Пятый раздел является вспомогательным. В нем проводятся дополнительные доказательства, что динамика частиц в плазме характеризуется хаотической динамикой. Наконец в шестом разделе показано, что плазма, которая находится в магнитном поле, может устойчиво существовать, если ее плотность не превосходит некоторое критическое значение. Лишние частицы будут удаляться из ансамбля. В заключении сформулированы основные результаты.

## 2. ДИНАМИКА АНСАМБЛЯ ЛИНЕЙНЫХ ОСЦИЛЛЯТОРОВ

Пусть у нас имеется система с гамильтонианом:

$$H = \sum_{i=0}^{N}\left(\frac{p_i^2}{2} + \omega_0^2 \frac{q_i^2}{2}\right) + \mu \cdot q_0 \cdot \sum_{j=1}^{N} q_j \qquad (1)$$

Эта система представляет собой $N$ связанных линейных осцилляторов. Гамильтониан (1) соответствует системе уравнений для описания динамики связанных линейных осцилляторов:

$$\ddot{q}_i + \omega_0^2 q_i = -\mu \cdot q_0$$
$$\ddot{q}_0 + \omega_0^2 q_0 = -\mu \cdot \sum_{i=1}^{N} q_i \qquad (2)$$

Причем для простоты мы рассматриваем систему, в которой все осцилляторы связаны друг с другом только через нулевой осциллятор (смотри рисунок 1). Нормальные частоты такой системы легко найти. Для этого решение системы (2) будем искать в виде:

$$q_i = a_i \exp(i \cdot \omega \cdot t), \ a_i = const \qquad (3)$$

Подставляя это решение в (2), легко получаем дисперсионное уравнение:

$$\left(-\omega^2 + \omega_0^2\right)^2 = \mu^2 N. \qquad (4)$$

Уравнение (4) дает следующие выражения для нормальных частот:

$$\omega = \pm \omega_0 \sqrt{1 \pm \mu \cdot \sqrt{N}/\omega_0^2}. \qquad (5)$$

Знаки + и – в формуле (5) перед корнем и под корнем независимы. Видно, что даже при очень маленьком коэффициенте связи, но при большом числе осцилляторов, одна из нормальных частот может быть очень маленькой (для случая знака (–) под корнем). Если выполняется неравенство:

$$\mu \cdot \sqrt{N} > \omega_0^2 \qquad (6)$$

то такой ансамбль существовать не может. Он разрушается. Численный анализ динамики системы (2) полностью подтверждает этот результат. Так, например, если десять осцилляторов в начальный момент времени расположены случайным образом в окрестности дна потенциальной ямы (смотри рис. 3) и коэффициент связи меньше 0.3, то колебания осцилляторов ограничены. Однако если мы слегка увеличим коэффициент связи ($\mu = 0.3334$), то динамика становится неустойчивой. Выполняется критерий разрушения (6). Ансамбль разрушается (смотри рис.4).

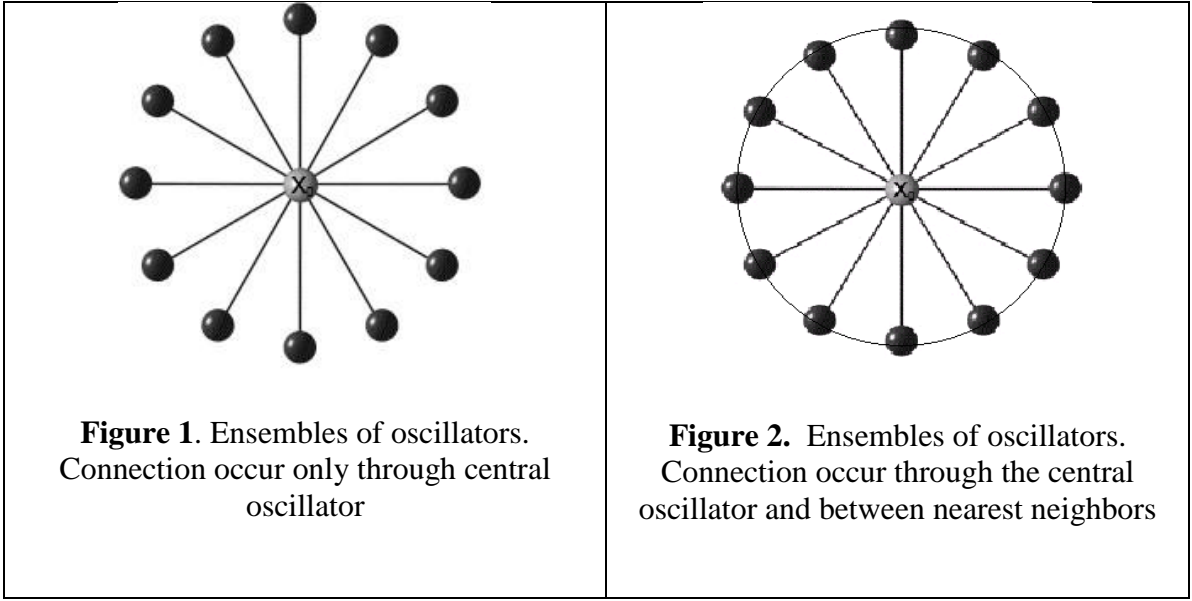

**Figure 1**. Ensembles of oscillators. Connection occur only through central oscillator

**Figure 2.** Ensembles of oscillators. Connection occur through the central oscillator and between nearest neighbors

Рассмотренный выше ансамбль является самым простым, модельным. Более реалистичным представляется ансамбль осцилляторов представленных на рисунке 2. В этом ансамбле, как и в предыдущем, все осцилляторы связаны с центральным осциллятором, а также связаны с ближайшими своими соседями. Кроме того частота центрального осциллятора отличается от частоты остальных осцилляторов. Система уравнений, которая описывает динамику такого ансамбля имеет следующий вид:

$$\ddot{q}_i + \omega_0^2 q_i = -\mu \cdot q_0 - \mu_1 (q_{i+1} + q_{i-1})$$
$$\ddot{q}_0 + \omega_1^2 q_0 = -\mu \cdot \sum_{i=1}^{N} q_i \qquad (7)$$

Для нахождения условий существования такого ансамбля систему (7) удобно переписать в виде:

$$\ddot{Q} + \omega^2 q_i = -\mu \cdot N q_0$$
$$\ddot{q}_0 + \omega_1^2 q_0 = -\mu \cdot Q \ . \qquad (8)$$

Where $\omega^2 = \omega_0^2 + 2\mu_1; \quad Q = \sum_{i=1}^{N} q_i$.

Будем искать решения системы (8) в виде: $Q \sim \exp(i \cdot \omega \cdot t)$. Тогда для нормальных частот найдем следующее выражение:

$$\omega^2 = \frac{1}{2}\left(\omega_1^2 + \omega_2^2\right)\left[1 \pm \sqrt{1 - 4\left(\omega_1^2 \cdot \omega_2^2 - \mu^2 N\right)/\left(\omega_1^2 + \omega_2^2\right)^2}\right], \qquad (9)$$

где - $\omega_2^2 = \omega_0^2 + 2\mu$ .

Все особенности динамики такой системы осцилляторов аналогичны особенностям предыдущей системы осцилляторов. Отметим, что условием разрушения этой системы осцилляторов будет условие:

$$\mu^2 N > \omega_2^2 \omega_1^2 . \qquad (10)$$

## 3. Линейные осцилляторы. Все связаны со всеми

Представляет интерес рассмотреть динамику наиболее часто встречающейся системы линейных осцилляторов, в которой каждый из осцилляторов связан со всеми другими осцилляторами. Систему уравнений, которая описывает динамику таких осцилляторов, можно записать:

$$\ddot{x}_i + \omega_0^2 x_i = -\mu \cdot \sum_{j \neq i}^{N} x_j = -\mu \cdot Q + \mu \cdot x_i . \qquad (11)$$

Введя новую функцию $Q = \sum_{i=0}^{N} x_i$, систему уравнений (11) можно переписать в более удобном виде:

$$\ddot{Q} + \omega_0^2 Q = -\mu \cdot (N-1) Q . \qquad (12)$$

Из уравнения (12) следует, что при $\mu > 0$, динамика системы осцилляторов остается колебательной. Однако при $\mu < 0$ и большом количестве осцилляторов ($N > 1 + \omega_0^2 / |\mu|$) - возникает неустойчивость. Например, пусть $N = 10; \omega_0 = 1$, тогда, если $|\mu| < 0.1$, то динамика регулярна и колебательная. Но уже при $|\mu| \geq 0.2$ ансамбль неустойчив. Численные расчеты эти выводы полностью подтверждают.

## 4. Ансамбль нелинейных осцилляторов

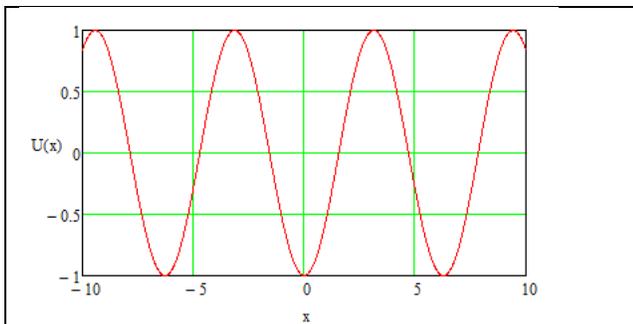

Figure 3. Вид потенциала, в котором находятся частицы

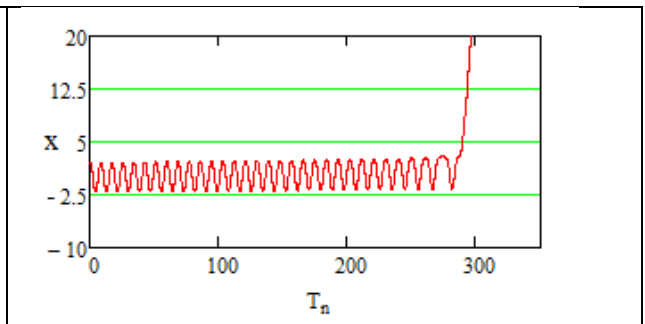

Figure 4. Характерная динамика частиц, которые покидают ансамбль взаимодействующих частиц. Причем эта динамика характерна как для линейных, так и для нелинейных осцилляторов.

Запишем систему дифференциальных уравнений, которая описывает динамику ансамбля математических маятников, в котором каждый осциллятор связан с каждым:

$$\ddot{x}_i + \sin x_i = -\mu \cdot \sum_{j \neq i}^{N} x_j \qquad i,j \in \{0,1,2....N\} \tag{13}$$

Система (13) является обобщением системы (11) на случай, когда осцилляторы нелинейные.

Система уравнения (13) исследовалась численными методами. Коэффициент связи между осцилляторами выбирался маленьким ($\mu < 10^{-3}$). В начальный момент времени частицы располагались в окрестности положения равновесия заряженных частиц. В общем случае в течение некоторого интервала времени частицы колеблются в потенциале. Они захвачены этим потенциалом. Однако в зависимости от величины коэффициента связи, начального положения частиц в потенциале и от их количества некоторые из частиц выбрасываются из потенциала. Вылет одной из частиц представлен на рисунке 4. Этот случай соответствует динамике десяти осцилляторам, которые связаны коэффициентом связи $\mu = 0.0005$. Максимальная безразмерная начальная скорость одной из частиц (не обязательно той, которая вылетает) равнялась 0.4. Следует отметить, что время вылета частицы из потенциала очень чувствительно к небольшим изменениям самого потенциала, коэффициентов связи, а также к положению частиц. К этому можно добавить, что динамика всех частиц локально неустойчива. Развивается динамический хаос. Если удалить частицы, которые вылетели из ансамбля, то динамика остальных частиц остается ограниченной.

### 5. Хаотическая динамика частиц в плазме

Движение частиц в плазме в общем случае подчиняются случайной динамике. Этот факт отмечается при описании результатов многочисленных теоретических и экспериментальных работ (см., например [11,12],). Строгое доказательство этого факта и критерии возникновения режимов с динамическим хаосом имеется для циклотронных резонансов (смотри [13-16]). Динамика частиц в полях многочисленных волн, которые возбуждаются в плазме, также должна быть хаотичной (в отсутствии циклотронных резонансов). Интуитивно этот факт не вызывает сомнения. Однако строгого доказательства этого факта и условий возникновения такой динамики, по-видимому, нет. Ниже, на простом примере показано, что в достаточно общем случае этот факт легко достаточно строго доказать. Докажем его. Известно, что плазма, особенно плазма в магнитном поле, обладает богатым спектром собственных колебаний волн. Покажем, что частицы плазмы в этих, даже регулярных полях, движутся хаотически. Для доказательства рассмотрим движение заряженных частиц в поле волнового пакета:

$$\ddot{z} = \frac{e}{m} \sum_i E_i \sin(k_i z - \omega_i t). \tag{14}$$

Рассмотрим вначале динамику в поле одной волны. Для такой динамики из уравнения (14) можно получить известный интеграл:

$$\frac{\dot{\varphi}^2}{2} - \Omega^2 \cos\varphi = H = const. \qquad (15)$$

Здесь: $\varphi = kz - \omega t$ $\quad \Omega^2 = \frac{|e|Ek}{m\omega^2} \quad \dot{\varphi} = d\varphi/d\tau; \ \tau = \omega t$.

Используя интеграл (15), находим ширину нелинейного резонанса:

$$\dot{\varphi}_{max} = +2\Omega, \quad \dot{\varphi}_{min} = -2\Omega. \qquad (16)$$

Для определения расстояния между резонансами обратим внимание, что эффективное взаимодействие частиц с волнами пакета происходит в условиях черенковского резонанса. В этом случае легко определить расстояние между резонансами:

$$\Delta\dot{\varphi} = -\Delta k \left[ v_0 - \frac{\Delta\omega}{\Delta k} k_0 \right]. \qquad (17)$$

При получении (17) учтено, что $v = v_{ph} = \omega/k$.

Используя выражения (16) и (17), находим условия возникновения локальной неустойчивости:

$$K = \left(\frac{\omega}{\Delta\omega}\right) \frac{2\Omega}{\left[1 - v_g/v_{ph}\right]} = N \frac{2\Omega}{\left[1 - v_g/v_{ph}\right]}. \quad K > 1 \qquad (18)$$

Здесь $v_g$ - групповая скорость; N - количество волн в пакете.

Анализируя формулы (17) и (18), можно сделать несколько важных заключений. Первое – видно (из формулы (17)), что если групповая скорость стремится к фазовой скорости волны, то расстояние между резонансами стремится к нулю. Это означает, что все волны пакета расположены на прямолинейном участке дисперсии. В фазовом пространстве резонансы таких волн все совпадают. Для частиц такие резонансы практически неразличимы. Динамика должна быть регулярной. Второе – если групповая скорость волн стремится к нулю (например, ленгмюровские волны в плазме), то, как видно из формулы (18), $1 < K \ll N; \Omega \ll 1$. В этом случае, как это впервые, по-видимому, было отмечено в работе [17], динамика частиц должна быть хаотичной. При этом отметим, что нерелятивистская динамика частиц практически всегда соответствует случаю $\Omega \ll 1$.

К полученному результату можно добавить дополнительные аргументы в пользу того, что динамика частиц в плазме хаотична. Покажем, что хаотическая динамика частиц может существенно облегчить их уход из удерживающего потенциала. Пусть, например, частицы движутся, как и в предыдущем разделе, в потенциале, представленном на рисунке 3. Динамика частицы в таком потенциале описывается уравнением математического маятника (13).

Смещение каждого из этих осцилляторов можно представить в виде:

$$x_i = \bar{x} + \delta_i, \tag{19}$$

где $\bar{x} = \left(\sum_{i=0}^{N} x_i\right) / N$ - средняя координата смещения нелинейного осциллятора; $\delta_i$ - случайное отклонение с нормальным распределением. Причем, $\langle \delta_i \rangle = 0$.

Для ансамбля осцилляторов величина $\bar{x}$ соответствует "центру инерции" ансамбля. В этом случае средние величины по ансамблю от функции $\sin x$ удобно представить в виде ряда по моментам:

$$\langle \sin(x_i) \rangle = \left[1 - \sum_{m=1}^{\infty} \frac{M_{2m}}{(2m!)}\right] \sin \bar{x}, \tag{20}$$

где $M_n = \langle (\delta)^n \rangle$ - моменты.

$$\ddot{\bar{x}} + \left[1 - \sum_{m=1}^{\infty} \frac{M_{2m}}{(2m!)}\right] \sin \bar{x} = 0 \tag{21}$$

При получении уравнений (20) и (21) предполагалось, что случайная динамика осцилляторов определяется законом распределения, близким к нормальному. По этой причине нечетные моменты в этих уравнениях не учитывались. Формула (21) справедлива и для описания хаотической динамики одного нелинейного осциллятора. В этом случае величина $\bar{x}$ соответствует среднему смещению осциллятора по реализациям.

Из этих уравнений следует, что область локализации частиц с ростом моментов становится меньшей. Эта область определяется шириной нелинейного резонанса:

$$\Delta = 4 \cdot \sqrt{\left[1 - \sum_{m=1}^{\infty} M_{2m} / (2m!)\right]}.$$

В результате глубина потенциальной ямы, в которой движутся частицы, также уменьшается. Поэтому даже небольшие по величине внешние силы легко выбрасывают частицы из области захвата.

## 6. Динамика частиц плазмы в магнитном поле.

Рассмотрим движение частиц с зарядом $e$ во внешнем магнитном поле, направленном вдоль оси $z$: $\vec{H}_0 = \{0, 0, H\}$. В таком поле частицы вращаются вокруг силовых линий магнитного поля. Они движутся с ускорением и излучают. В плазме такие частицы взаимодействуют друг с другом. Будем считать плазму идеальной, поэтому кулоновское взаимодействие частиц учитывать не будем. Будем считать, что взаимодействие осуществляется с помощью полей, которые частицы излучают при вращении. Напряженность электрического поля в окрестности другой частицы будет:

$$\vec{E} = -\frac{e \cdot \dot{\vec{v}}}{c^2 R} \tag{22}$$

Здесь $e$ - заряд частицы, $\dot{\vec{v}}$ - ускорение частицы при ее вращении, $c$ - скорость света, $R$ - расстояние между частицами.

Рассмотрим вначале взаимодействие двух частиц. Будем рассматривать только поперечное относительно магнитного поля движение частиц. Динамику одной из частиц можно описать следующим уравнением:

$$m\ddot{\vec{r}} = \frac{e}{c}\left[\dot{\vec{r}}\vec{H}_0\right] - \frac{e^2\dot{\vec{v}}}{c^2 R}; \qquad (23)$$

Это векторное уравнение удобно расписать в виде системы уравнений для компонент скоростей первой частицы:

$$\dot{v}_{x1} = \omega_H v_{y1} - \mu \cdot \dot{v}_{x2} \qquad (24)$$

$$\dot{v}_{y1} = -\omega_H \cdot v_{x1} - \mu \cdot \dot{v}_{y2}$$

Где $\omega_H = \frac{eH}{mc}$ - круговая частота вращения частицы в магнитном поле, $\mu = \frac{e^2}{Rmc^2}$ влияние второй частицы на динамику первой частицы (коэффициент связи между частицами).

Дифференцируя систему уравнений (24) и учитывая саму систему (24), получим следующую систему уравнений:

$$(\ddot{v}_{x1} + \omega_H^2 \cdot v_{x1}) = 2\mu\omega_H^2 v_{x2}$$
$$\ddot{v}_{y1} + \omega_H^2 v_{y1} = 2\mu\omega_H^2 v_{y2} \qquad (25)$$

Аналогичные системы уравнений можно записать и для динамики второй частицы. Если частиц много, то уравнения, которые описывают динамику компонент скоростей, можно представить в виде:

$$\ddot{v}_k + v_k = 2\sum_{j \neq k} \mu_j v_j \qquad (26)$$

В уравнении (26) зависимая переменная $v_k$ определяет либо $x$, либо $y$ компоненту скорости $k$-ой частицы. Кроме того, введено новое время $\tau = \omega_H t$. Коэффициенты связи $\mu_j$, которые стоят в правой части уравнения (26) под знаком сумы, отличаются друг от друга только расстоянием между частицами $R_j$. Обратим внимание, что при большом количестве осцилляторов ($N >> 1$) и малых коэффициентах связи ($\mu_j << 1$) правая часть уравнений (26) для всех осцилляторов (для всех $k$) одинакова. В этом случае уравнения (26) можно существенно упростить:

$$\ddot{v}_k + \left(1 - 2\sum_{j=1}^{N}\mu_j\right)v_k = 0 \qquad (27)$$

Условие неустойчивости приобретает вид:

$$2\sum_{j=1}^{N}\mu_j > 1 \tag{28}$$

Условие (28) можно переписать для плотности частиц. Действительно, полное число взаимодействующих частиц равно плотности частиц умноженной на объем, который занимают взаимодействующие частицы ($N = nV$). Объем шарового слоя радиуса $r$ и толщиной $dr$ равен $V_{laer} = 4\pi r^2 dr$ Используя эти данные левую часть неравенства (28) можно переписать:

$$2\sum_{j=1}^{N}\mu_j = 8\pi n \int_{0}^{R_{max}} \frac{r^2 dr}{r} \frac{e^2}{mc^2} = 4\pi \frac{e^2}{mc^2} n R_{max}^2 = \mu N$$

А условие (28) приобретет вид:

$$n > 3 \cdot 10^{11} / R_{max}^2 \tag{29}$$

### 7. Заключение.

Сформулируем наиболее важные результаты:

1. Ансамбль заряженных частиц (электронов) в магнитном поле может устойчиво существовать, только если плотность этих частиц меньше некоторого критического значения (см. формулу (29)). При большем числе частиц «лишние» частицы удаляются из ансамбля (см. разделы 2-4). Развитие этой неустойчивости означает, что плазма в данных моделях, например плазменный цилиндр, при плотностях больше $10^{12}$ начинает одеваться высокоэнергичной электронной шубой. Плазменный стержень будет положительно заряжен. Возникает удерживающий потенциал. Динамика частиц в этой шубе хаотична.
2. Хаотическая динамика частиц в потенциалах, которые удерживают частицы, может существенно ослабить эффективность их удержания (см. формулу (21)).
3. Выше, основное внимание было уделено условиям разрушения колебательной динамики системы осцилляторов. Однако значительный интерес могут представить и устойчивые ансамбли при большом числе осцилляторов. Примеры таких ансамблей приведены в разделе 2. При большом числе осцилляторов, как видно из формулы (5), нормальная частота ансамбля может быть существенно ниже, чем парциальные частоты отдельных осцилляторов $\omega << \omega_0$. При этом, если в какой-то момент времени основной осциллятор, который связывает все осцилляторы в единый ансамбль, исчезает, то осцилляторы колеблются со своей парциальной частотой. Такую особенность можно использовать. Например, если в качестве основного связывающего осциллятора рассматривать модулированный электронный пучок конечной длительности ($0 < t < T$), то по истечении времени $T$ он исчезает. Если пучок будет промодулирован на нормальной частоте ансамбля (on $\omega$), то он резонансно возбудит колебания всего ансамбля. После своего исчезновения отдельные осцилляторы будут колебаться на своих парциальных частотах $\omega_0$ (других собственных частот нет). Такой механизм позволяет эффективно преобразовать энергию низкочастотного колебания (энергию модулированного пучка) в энергию высокочастотных колебаний. Этот механизм может быть использован для создания новых типов источников излучения в плохо освоенных диапазонах частот. Например в терагерцовом диапазоне.

4. Возникает вопрос о том, какие системы осцилляторов будут устойчивыми, т.е. динамика их останется колебательной вне зависимости от числа осцилляторов. В частности, в монографиях [8,9] рассмотрены системы осцилляторов, частота которых, а также коэффициенты связи между осцилляторами полностью зависят от характеристик потенциала (определяются вторыми частными производными от потенциала), в котором находятся заряженные частицы (осцилляторы). Причем каждый осциллятор связан со всеми другими осцилляторами, а коэффициенты связей являются взаимными – они симметричны по своим индексам $\mu_{ik} = \mu_{ki}$. В этом случае такие системы осцилляторов обладают только колебательными модами. Неустойчивость в таких ансамблях не развивается. Полученные выше результаты позволяют выделить две группы осцилляторов, в которых возможно развитие неустойчивостей. К первой группе можно отнести ансамбли осцилляторов, в которых не все осцилляторы связаны со всеми другими. Примерами являются системы осцилляторов, рассмотренные во втором разделе. Такие системы осцилляторов легко реализовать искусственно. В плазме, в естественных условиях, они не существуют. Однако в плазме, особенно в плазме, удерживаемой магнитным полем, можно выделить вторую группу неустойчивых ансамблей осцилляторов. Связи между осцилляторами в этой группе определяются полями, которые возбуждаются самими осцилляторами. Плазма диамагнитна, поэтому такие поля препятствуют процессу, который их возбудил. Результатом такой физической особенности является изменение знака коэффициентов связи. При достаточно большом числе осцилляторов такие системы, как мы видели выше, становятся неустойчивыми. Следует также иметь в виду, что развитие хаотических режимов способствует освобождению частиц из области удерживающего потенциала.

**5.** Описанные выше модели максимально упрощены. В экспериментальных условиях, например, магнитное поле является неоднородным. При этом парциальные частоты осцилляторов становятся различными. Эффективность взаимодействия осцилляторов с различными частотами значительно меньше, чем эффективность взаимодействия одинаковых осцилляторов. Поэтому можно ожидать, что необходимое число частиц для разрушения ансамбля будет большим, чем, например, определяет формула (28). При этом, однако, тенденция к появлению нормальных мод, частота которых значительно меньше парциальных частот отдельных осцилляторов, очевидно, будет сохраняться. В этом случае может возникнуть коллективная низкочастотная мода, которая будет выбрасывать частицы плазмы на стенку. Заметим, что анализ таких систем затруднен даже численными методами.

## References

18. Nicholas A. Krall, Alvin W. Trivelpiece, Principles of Plasma Physics, McGRAW-HILL Book Company .1973, 526
19. Chen, F.F.; *Introduction to Plasma Physics and Controlled Fusion.* 1984, l,2 nded. (Plenum Press,).
20. Fusion physics. Edited by: Mitsuru Kikuchi; Karl Lackner; Minh Quang. *Trans International Atomic Energy* Agency, Vienna, 2012.
21. Greenwald, M., Density limits in toroidal plasmas, *Plasma Phys. Control. Fusion,* 2002, 44 , R27-R53.

22. Puiatti, M.E.; Scarin, P.; Spizzo, G.; et al.. High density limit in reversed field pinches. *Phys. Plasmas.* 2009, 16, 012505
23. Gates, D.A.; Delgado-Aparicio, L. Origin of Tokamak Density Limit Scalings, *Phys. Rev. Lett.* 2012, 108,165004.
24. Baek, S.G.; Wallace, G.M.; Bonoli, P.T.; Brunner, D.; Faust, I.C.; Hubbard, ; Hughes, J.W. ;. LaBombard, B.; Parker, R.R; Porkolab, M.; Shiraiwa, S.; and Wukitch, S. Observation of Efficient Lower Hybrid Current Drive at High Density in Diverted Plasmas on the Alcator C-Mod Tokamak. *Phys. Rev. Lett.* 121, 055001 – Published 3 August 2018.
25. Landau, L. D.; Lifshitz, E.M. Mechanics, *Elsevier Butterworth Heinemann,* 172.
26. Magnus, K. Oscillations. *Moscow "Mir"*. 1982, 304 (In Russia).
27. Kuklin, V.M.; Litvinov, D.N.; Sevidov, S.M.; Sporov, A.E. Simulation of synchronization of nonlinear oscillators by the external field. *East European Journal of Physics*. 2017, 4, (1), 75-84.
28. Shustin, E.G.; Isaev, N.V.; Temiryazeva, M.P.; Tarakanov, V.P.; Fedorov, Yu.V. Beam-Plasma Discharge in a Weak Magnetic Field as a Source of Plasma For a Plasma- Chemical Reactor
*Kharkov. Problems of Atomic Science and Technology. Series: Plasma electronics and new acceleration methods.* 2008, № 4 (6), 169.

29. Meshkov, I.N.; Nagaitsev, S.S.; Seleznev, I.A.; Syresin, E.M. Beam – Plasma Discharge at Electron Beam Injection Into Rare Gas. Preprint 90-12. *Novosibirsk Institute of Nuclear Physics.* 1990, 630090, 15.
30. Moiseev, S.S.; Buts, V.A.; Erokhin, N.S. Peculiarities of Charged Particle Dynamics under Cyclotron Resonance Conditions. *Plasma Physics Reports*. 2016, 42, N 8, 761-768.
31. Buts, V. A.; Kuzmin, V. V.; Tolstoluzhsky, A. P. Features of the dynamics of particles and fields at cyclotron resonances. *ZhETF,* 2017, 152, 4 (10), 767-780.
32. Buts, V. A.; Lebedev A. N.; and Kurilko V. I., The Theory of Coherent Radiation by Intense Electron Beams. *Springer,* Berlin, 2006.
33. Buts, V.A. Regular and chaotic dynamics of charged particles during wave-particle interactions. *Problems of theoretical physics. Series. Problems of theoretical and mathematical physics.* 2017, Kharkiv, 2, 122-241.
34. Zaslavsky, G.M.; Chirikov, B.V.; Stochastic instability of nonlinear oscillations. *Phys. Usp. 1971*, 105, 1-42.